# Thermodynamic perturbation theory for self assembling mixtures of multi - patch colloids and colloids with spherically symmetric attractions

Bennett D. Marshall[1] and Walter G. Chapman
Department of Chemical and Biomolecular Engineering
Rice University
6100 S. Main
Houston, Texas  77005

## Abstract

In this paper we extend our previous theory [B. D. Marshall and W.G. Chapman, J. Chem. Phys. 139, 104904 (2013)] for mixtures of single patch colloids (*p* colloids) and colloids with spherically symmetric attractions (*s* colloids) to the case that the *p* colloids can have multiple patches. The theory is then applied to the case of a binary mixture of bi-functional *p* colloids which have an *A* and *B* type patch and *s* colloids which are not attracted to other *s* colloids and are attracted to only patch *A* on the *p* colloids. This mixture reversibly self assembles into both colloidal star molecules, where the *s* colloid is the articulation segment and the *p* colloids form the arms, and free chains composed of only *p* colloids.  It is shown that temperature, density, composition and relative attractive strengths can all be varied to manipulate the number of arms per colloidal star molecule, average arm length and the fraction of chains (free chains + arms) which are star arms.

---

[1] Author to whom correspondence should be addressed
 Email:  bennettd1980@gmail.com

# I. Introduction

In recent years there has been a rapid increase in our ability to synthesize patchy colloids.[1] Patchy colloids are colloidal particles which have discrete attractive patches located on their surface. This results in anisotropic orientation dependent attractive potentials between the colloids and encodes the colloids with a defined valence. The valence and anisotropic potentials can be manipulated by varying the number[2], size[3-7], shape[8], interaction range[9] and relative location[10, 11] of attractive patches on the surface of the colloid. Through the tailored design of these patch parameters these colloids can be programmed to self assemble into pre-determined structures such as colloidal molecules[12], the Kagome lattice[13] and diamond structures for photonic applications[14]; as well as form fluid phases such as gels[15-17], empty liquids[2, 18] and fluids which exhibit reentrant phase behavior[19]. These patchy colloids have been synthesized by glancing angle deposition[20, 21], the polymer swelling method[12] and by stamping the colloids with patches of single stranded DNA.[22] With the rapid advance in our synthetic ability it seems certain that, in the near future, synthetic chemist will be able to produce these patchy colloids to a high degree of precision for specific design specifications.

Recently[22], researchers synthesized mixtures of patchy *p* and spherically symmetric *s* colloids by binding DNA to the surfaces of the colloids. The *p* colloids had a single sticky patch terminated with type *A* single stranded DNA sticky ends and the *s* colloids were uniformly coated with DNA terminated with complementary type *B* single stranded DNA sticky ends. The DNA types were chosen such that there were *AB* attractions but no *AA* or *BB* attractions. That is, *s* colloids attract the anisotropic *p* colloids, but *s* colloids do not attract other *s* colloids and *p* colloids do not attract other *p* colloids. It was shown that this mixture would reversibly self assemble into clusters where a single *s* colloid would be bonded to some number *n* of *p* colloids (colloidal star molecules consisting of *n* arms of length one).

Subsequently, Marshall and Chapman[23] developed a perturbation theory to describe the thermodynamics and self assembly of this mixture using the two density statistical mechanics of Wertheim[24, 25]. The challenge of developing such a theory is the inherent higher order nature of the attraction between the patches on the *p* colloids and the *s* colloids. Wertheim's two density formalism takes on a simple form if it can be assumed that each patch is only singly bondable. This is the single bonding condition. To allow for a patch to bond *n* times, $n + 1$ body correlations must be accounted for. If the single bonding condition holds, only pair correlations are required. For the *p* and *s* colloid mixture considered[23], the *s* colloid, which has spherically symmetric attractions, could bond to a maximum of thirteen *p* colloids. Clearly, this is well beyond the single bonding condition. Through the introduction of cluster partition functions, Marshall and Chapman[23] where able to develop a relatively simple theory to model this mixture. By comparison to simulation data, this new theory was shown to be very accurate for the predication of the reversible and temperature dependent self assembly of the system into colloidal star molecules and the resulting thermodynamic properties.

The theory discussed above is limited by the fact that the *p* colloids only have a single patch. What about the many patch case? For instance, consider the case (case I) where the *p* colloids have two patches (an *A* patch and *B* patch located on opposite poles of the colloid) and the *s* colloids are only attracted to the type *A* patch on the *p* colloid. As before, there are *AB* attractions but no *AA* or *BB* attractions. The simple addition of this extra patch on the *p* colloid significantly increases the complexity and richness of the behavior of the *s* and *p* colloid mixture, as compared to the one patch case. Now there will be a competition between the *p* colloids associating into free chains of colloids and the *p* colloids bonding to the *s* colloids to form colloidal star molecules. Unlike the one patch case, now the arm number distribution of colloidal star molecules as well as arm length distribution will vary with temperature, composition and density. Since there are no *BB* attractions, no two *s* colloids can be connected by a path of association bonds (two arms on separate star molecules cannot form an association bond). This leads us into case II. Case II is identical to case I with the only change being we allow *BB* attractions. For this case the *s* colloids will now act as junctions which seed network formation. For low *s* colloid mole fraction this mixture would likely form an empty liquid[2] as the distance between *s* colloid junctions increases.

These *s* and *p* colloid mixtures represent a new class of materials with an enormous potential. In this paper we develop a simple perturbation theory to model mixtures of this type. We will develop the theory in Wertheim's multi – density formalism for multi-site associating fluids.[26] The anisotropic attractions between *p* colloids will be treated in standard first order perturbation theory (TPT1)[27-29] while the attractions between *p* and *s* colloids are treated in a modification of the results of our previous paper[23]. The *p* colloids are allowed to have any number of patches and throughout the paper we consider a hard sphere reference system. To include attractions between *s* colloids one simply needs to use an appropriate reference system.

Once developed, we apply the theory to case I discussed above; case II will be considered in a future paper. We show that this *s* and *p* colloid mixture self assembles into a mixture of free colloidal chains consisting of only *p* colloids and colloidal star molecules which consist of an *s* colloid articulation segment and *p* colloid arms. We show that the fraction of chains which are star arms, average number of star arms per *s* colloid, average length of star arms, fraction of *p* colloids bonded *k* times and resulting thermodynamics can be manipulated by varying temperature, density, composition relative attractions. As a quantitative test, we compare theoretical predictions to new monte carlo simulation results. The theory is shown to be accurate.

In section II the new theory will be developed for the general case in which the *p* colloid is allowed to have an arbitrary number of patches. In section III we specialize the theory to the two patch case I discussed above and consider the low temperature limit of this model. In section IV we will discuss the simulation methodology used and in section V we study the self assembly of the case I mixture in detail. Finally, in section VI we give conclusions and discuss other theoretical approaches to model mixtures of this type.

## II. Theory

In this section we derive the theory for a two component mixture of patchy $p$ and spherically symmetric $s$ colloids of diameter $d$. We choose the case of equal diameters for notational simplicity, extension of the general approach to mixtures of different diameters is straight forward. The $p$ colloids have a set $\Gamma^{(p)} = \{A, B, C...\}$ of short range attractive patches whose sizes are determined by the critical angle $\beta_c^{(L)}$ which defines the solid angle of patch $L$ as $2\pi(1-\cos\beta_c^{(L)})$. The $s$ colloids are thought of as a colloid with a single large patch of critical angle $\beta_c^{(s)} = 180°$. A diagram of these types of colloids can be found in Fig. 1 for the case that attractions are mediated by grafted single stranded DNA with sticky ends. This is a generalization of the one patch case studied theoretically in our previous paper[23] and experimentally by Feng et al.[22]

The potential of interaction between two $p$ colloids is given by the sum of a hard sphere potential $\phi_{HS}^{(p,p)}(r_{12})$ and orientation dependant attractive patchy potential

$$\phi^{(p,p)}(12) = \phi_{HS}^{(p,p)}(r_{12}) + \sum_{A\in\Gamma^{(p)}}\sum_{B\in\Gamma^{(p)}} \phi_{AB}^{(p,p)}(12) \qquad (1)$$

The notation $(1) \equiv (\vec{r}_1, \Omega_1)$ represents the position $\vec{r}_1$ and orientation $\Omega_1$ of colloid 1 and $r_{12}$ is the distance between the colloids. Here we follow Kern and Frenkel[30] who employed a potential for conical association sites[27, 31] to model the patchy attractions

$$\phi_{AB}^{(p,p)}(12) = \begin{cases} -\varepsilon_{AB}^{(p,p)}, & r_{12} \leq r_c \text{ and } \beta_A \leq \beta_c^{(A)} \text{ and } \beta_B \leq \beta_c^{(B)} \\ 0 & otherwise \end{cases} \qquad (2)$$

which states that if colloids 1 and 2 are within a distance $r_c$ of each other and each colloid is oriented such that the angles between the site orientation vectors and the vector connecting the two segments, $\beta_A$ for colloid 1 and $\beta_B$ for colloid 2, are both less than the critical angle, $\beta_c^{(A)}$ for $A$ and $\beta_c^{(B)}$ for $B$, the two sites are considered bonded and the energy of the system is decreased by a factor $\varepsilon_{AB}^{(p,p)}$.

The interaction between $p$ and spherically symmetric $s$ colloids is similarly defined as

$$\phi^{(s,p)}(12) = \phi_{HS}^{(s,p)}(r_{12}) + \sum_{A \in \Gamma^{(p)}} \phi_A^{(s,p)}(12) \quad (3)$$

Where the attractive potential is given by[23]

$$\phi_A^{(s,p)}(12) = \begin{cases} -\varepsilon_A^{(s,p)}, & r_{12} \leq r_c \text{ and } \beta_A \leq \beta_c^{(A)} \\ 0 & otherwise \end{cases} \quad (4)$$

which states that if $s$ colloid 1 and $p$ colloid 2 are within a distance $r_c$ of each other, and the $p$ colloid is oriented such that the angle between the site $A$ orientation vector and the vector connecting the two segments $\beta_A$ is less than the critical angle $\beta_c^{(A)}$, the two colloids are considered bonded and the energy of the system is decreased by a factor $\varepsilon_A^{(s,p)}$. Lastly the $s$ colloids are assumed to interact with hard sphere repulsions only, that is

$$\phi^{(s,s)}(12) = \phi_{HS}^{(s,s)}(r_{12}) \quad (5)$$

We develop the theory in the multi density formalism of Wertheim[26, 29, 32] where each bonding state of a colloid is assigned a number density. The density of species $k = \{p, s\}$ bonded at the set of patches $\alpha$ is given by $\rho_\alpha^{(k)}$. To aid in the reduction to irreducible graphs, Wertheim introduced the density parameters $\sigma_\gamma^{(k)}$

$$\sigma_\gamma^{(k)} = \sum_{\alpha \subset \gamma} \rho_\alpha^{(k)} \tag{6}$$

where the empty set $\alpha = \varnothing$ is included in the sum. Two notable cases of Eq. (6) are $\sigma_{\Gamma^{(k)}}^{(k)} = \rho^{(k)}$ and $\sigma_o^{(k)} = \rho_o^{(k)}$; where $\rho^{(k)}$ is the total number density of species $k$ and $\rho_o^{(k)}$ is the density of colloids $k$ not bonded at any patch (monomer density).

In Wertheim's multi – density formalism the change in free energy due to association is given by[26, 32, 33]

$$\frac{A^{AS}}{Vk_BT} = \sum_k \left( \rho^{(k)} \ln\left(\frac{\rho_o^{(k)}}{\rho^{(k)}}\right) + Q^{(k)} + \rho^{(k)} \right) - \Delta c^{(o)}/V \tag{7}$$

where $V$ is the system volume, $T$ is temperature and $Q^{(k)}$ is given by

$$Q^{(k)} = -\rho^{(k)} + \sum_{\substack{\gamma \subset \Gamma^{(k)} \\ \gamma \neq \varnothing}} c_\gamma^{(k)} \sigma_{\Gamma^{(k)}-\gamma}^{(k)} \tag{8}$$

The term $\Delta c^{(o)}$ is the associative contribution to the fundamental graph sum which encodes all association attractions between the colloids, and $c_\gamma^{(k)}$ is obtained from the relation

$$c_\gamma^{(k)} = \frac{\partial \Delta c^{(o)}/V}{\partial \sigma_{\Gamma^{(k)}-\gamma}^{(k)}} \tag{9}$$

where in Eq. (9) $\gamma \neq \varnothing$. The attractive fundamental graph sum $\Delta c^{(o)}$ is decomposed as

$$\Delta c^{(o)} = \Delta c_{pp}^{(o)} + \Delta c_{sp}^{(o)} \tag{10}$$

Where $\Delta c_{pp}^{(o)}$ accounts for the attractions between $p$ colloids and $\Delta c_{sp}^{(o)}$ accounts for the attraction between $p$ and $s$ colloids. We treat the interaction between $p$ colloids in first order perturbation theory (TPT1) giving $\Delta c_{pp}^{(o)}$ as[32]

$$\Delta c_{pp}^{(o)}/V = \frac{1}{2} \sum_{L \in \Gamma^{(p)}} \sum_{M \in \Gamma^{(p)}} \sigma_{\Gamma^{(p)}-L}^{(p)} \xi \kappa_{LM} f_{LM}^{(p,p)} \sigma_{\Gamma^{(p)}-M}^{(p)} \tag{11}$$

where $\kappa_{LM} = (1-\cos\beta_c^{(L)})(1-\cos\beta_c^{(M)})/4$ is the probability that two colloids are oriented such that patch $L$ on colloid 1 can bond to patch $M$ on colloid 2 and

$$\xi \approx 4\pi d^2 (r_c - d) y_{HS}(d) \tag{12}$$

is the integral of the hard sphere reference cavity correlation function over the bond volume. Lastly, the term $f_{LM}^{(p,p)} = \exp(\varepsilon_{LM}^{(p,p)}/k_BT) - 1$ is the magnitude of the association Mayer function.

The contribution $\Delta c_{sp}^{(o)}$ cannot be obtained in TPT1 due to the fact that we must move beyond Wertheim's single bonding condition[26] which restricts each patch to bonding only once. The $s$ colloids exhibit spherical symmetry meaning they can bond multiple times. The maximum number of bonds is simply the maximum number of $p$ colloids $n^{max}$ which can pack in the $s$ colloids bonding shell. We can rewrite $\Delta c_{sp}^{(o)}$ as

$$\Delta c_{sp}^{(o)} = \sum_{n=1}^{n^{max}} \Delta c_n \qquad (13)$$

where $\Delta c_n$ is the contribution for $n$ patchy colloids bonded to a single $s$ colloid. We approximate $\Delta c_n$ in a generalization Wertheim's single chain approximation[29, 32] and consider all graphs consisting of a single associated cluster with $n$ patchy colloids bonded to a $s$ colloid with $n$ association bonds. We then simplify the results as described in our previous paper[23] to obtain

$$\Delta c_n^{(o)}/V = \frac{1}{n!}\rho_o^{(s)}\Delta^n \delta^{(n)} \Xi^{(n)} \qquad (14)$$

where

$$\Delta = y_{HS}(d) \sum_{L \in \Gamma^{(p)}} \sigma_{\Gamma^{(p)}-L}^{(p)} f_L^{(s,p)} \sqrt{\kappa_{LL}} \qquad (15)$$

and $\delta^{(n)}$ is the second order correction to the first order superposition of the many body correlation function for the associated cluster. This term is evaluated using the branched TPT2 solution of Marshall and Chapman[34] as

$$\delta^{(n)} = \begin{cases} (1+4\lambda)^{\frac{n-3}{2}} \left( \dfrac{1+\sqrt{1+4\lambda}}{2} \right)^3 & for \quad n>1 \\ 1 & for \quad n=1 \end{cases} \qquad (16)$$

The term $\lambda = 0.2336\eta + 0.1067\eta^2$ where $\eta$ is the packing fraction.[35] Finally, the terms $\Xi^{(n)}$ are the cluster partition functions which are independent of temperature, density and composition and were evaluated in our previous paper.[23]

Now that $\Delta c^{(o)}$ has been fully specified the densities of the various bonding states can be calculated through the relation[26]

$$\frac{\rho_\gamma^{(k)}}{\rho_o^{(k)}} = \sum_{P(\gamma)=\{\tau\}} \prod_\tau c_\tau^{(k)} \qquad (17)$$

where $P(\gamma)$ is the partition of the set $\gamma$ into non-empty subsets. For example, the density $\rho_{ABC}$ is given by $\rho_{ABC} = \rho_o(c_{ABC} + c_{AB}c_C + c_{BC}c_A + c_{CA}c_B + c_A c_B c_C)$. Since the spherically symmetric colloid is a single patch, Wertheim's theory only assigns two densities; the density of $s$ colloids not bonded $\sigma_o^{(s)} = \rho_o^{(s)}$ and the density of $s$ colloids which are bonded $\rho_b^{(s)}$. The density $\rho_b^{(s)}$ is obtained from Eq. (17) as

$$\rho_b^{(s)} = \rho_o^{(s)} \frac{\partial \Delta c^{(o)}/V}{\partial \rho_o^{(s)}} = \sum_{n=1}^{n^{max}} \rho_n^{(s)} \qquad (18)$$

where $\rho_n^{(s)}$ is the density of $s$ colloids bonded $n$ times which we identify as

$$\rho_n^{(s)} = \frac{\Delta c_n}{V} \qquad for\ n > 0 \qquad (19)$$

Turning our attention to the $p$ colloids we note from Eqns. (9) and (10) that $c_\gamma^{(p)} = 0$ for $n(\gamma) > 1$ which results in the following rule from Eq. (17)

$$\frac{\rho_\gamma^{(k)}}{\rho_o^{(k)}} = \prod_{A \in \gamma} \frac{\rho_A^{(k)}}{\rho_o^{(k)}} \tag{20}$$

Equation (20) leads to the following relation for the fraction of colloids *not bonded* at patch $A$ $X_A^{(p)} = \sigma_{\Gamma^{(p)}-A}^{(p)} / \rho^{(p)}$ as

$$X_A^{(p)} = \frac{1}{1 + c_A^{(p)}} \tag{21}$$

as well as the relation for the monomer fraction

$$X_o^{(p)} = \prod_{A \in \Gamma^{(p)}} X_A^{(p)} \tag{22}$$

We obtain $c_A^{(p)}$ from Eqns. (9) and (10) as

$$c_A^{(p)} = \sum_{M \in \Gamma^{(p)}} \xi \kappa_{AM} f_{AM}^{(p,p)} \rho^{(p)} X_M^{(p)} + \sum_{n=1}^{n^{\max}} \frac{\rho_o^{(s)n}}{n!} \sqrt{\kappa_{AA}} y_{HS}(d) f_A^{(s,p)} \Delta^{n-1} \Xi^{(n)} \delta^{(n)} \tag{23}$$

Since $c_\gamma^{(p)} = 0$ for $n(\gamma) > 1$ we obtain from Eq. (8)

$$\frac{Q^{(p)}}{\rho^{(p)}} = -1 + \sum_{A \in \Gamma^{(p)}} c_A^{(p)} X_A^{(p)} \tag{24}$$

which when combined with Eq. (21) gives

$$\frac{Q^{(p)}}{\rho^{(p)}} = -1 + \sum_{A \in \Gamma^{(p)}} \left(1 - X_A^{(p)}\right) \tag{25}$$

For the spherically symmetric $s$ colloids Eq. (8) simply gives

$$\frac{Q^{(s)}}{\rho^{(s)}} = -X_o^{(s)} \tag{26}$$

We can now write $\Delta c_{pp}^{(o)}/V$ as

$$\frac{\Delta c_{pp}^{(o)}}{V} = \frac{1}{2} \sum_{A \in \Gamma^{(p)}} \rho^{(p)} X_A^{(p)} c_A^{(p)} - \sum_{n=1}^{n^{max}} \frac{n}{2} \rho^{(s)} X_n^{(s)} \tag{27}$$

The term $X_n^{(s)} = \rho_n^{(s)}/\rho^{(s)}$ is the fraction of spherically symmetric colloids bonded $n$ times obtained from Eq. (19). Combining these results we obtain the free energy

$$\frac{A^{AS}}{Nk_BT} = x^{(s)}\left(\ln X_o^{(s)} - X_o^{(s)} + 1 + \sum_{n=1}^{n^{max}}\left(\frac{n}{2} - 1\right)X_n^{(s)}\right) + \left(1 - x^{(s)}\right)\sum_{A \in \Gamma^{(p)}}\left(\ln X_A^{(p)} - \frac{X_A^{(p)}}{2} + \frac{1}{2}\right) \tag{28}$$

where $N$ is the total number of colloids in the system and $x^{(s)}$ is the mole fraction of $s$ colloids. Using the relation

$$\sum_{n=0}^{n^{max}} X_n^{(s)} = 1 \tag{29}$$

and the definition of the average number of bonds per $s$ colloid

$$\bar{n} = \sum_{n=0}^{n^{max}} n X_n^{(s)} \tag{30}$$

Eq. (28) can be further simplified as

$$\frac{A^{AS}}{Nk_BT} = x^{(s)}\left(\ln X_o^{(s)} + \frac{\bar{n}}{2}\right) + \left(1 - x^{(s)}\right) \sum_{A \in \Gamma^{(p)}} \left(\ln X_A^{(p)} - \frac{X_A^{(p)}}{2} + \frac{1}{2}\right) \tag{31}$$

Where the fractions $X_A^{(p)}$ are obtained by solving Eqns. (21) in conjunction with the relation

$$X_o^{(s)} = \frac{1}{1 + \sum_{n=1}^{n^{max}} \frac{1}{n!} \Delta^n \Xi^{(n)} \delta^{(n)}} \tag{32}$$

which was obtained using Eq. (29). With Eq. (32) we conclude this section. The chemical potential $\mu$, pressure $P$ and excess internal energy $E$ are all calculated in the appendix.

## III. Application to 2 patch colloids

Here we consider the case, case I discussed in the introduction, where the *p* colloid has an *A* type and *B* type patch where $\varepsilon_{AA}^{(p,p)} = \varepsilon_{BB}^{(p,p)} = 0$. For attractions between the *p* and *s* colloid we set

$$\varepsilon_A^{(s,p)} = C\varepsilon_{AB}^{(p,p)} \tag{33}$$

$$\varepsilon_B^{(s,p)} = 0$$

where the constant $C$ is defined by Eq. (33). The restrictions $\varepsilon_B^{(s,p)} = \varepsilon_{BB}^{(p,p)} = 0$ will suppress the formation of a network, meaning a branch emanating from an *s* colloid cannot terminate on another *s* colloid. This situation is depicted in Fig.2. This system represents a mixture of colloids which exhibits the reversible and temperature dependant self assembly into colloidal star molecules and free chains, where the *s* colloids are the articulation segments for the star molecules and the *p* colloids make up the arms of the star molecules and the segments of the free chains.

From Eqns. (23), (33) and (30)

$$c_A^{(p)} = \xi \kappa_{AB} f_{AB}^{(p,p)} \rho^{(p)} X_B^{(p)} + \frac{\rho^{(s)}}{\rho^{(p)}} \frac{\bar{n}}{X_A^{(p)}} \tag{34}$$

and

$$c_B^{(p)} = \xi \kappa_{AB} f_{AB}^{(p,p)} \rho^{(p)} X_A^{(p)} \tag{35}$$

Equations (34) – (35) combined with (21) give the fractions not bonded

$$X_B^{(p)} = \frac{1}{1 + \xi \kappa_{AB} f_{AB}^{(p,p)} \rho^{(p)} X_A^{(p)}} \tag{36}$$

and

$$X_A^{(p)} = \frac{1}{1 + \xi \kappa_{AB} f_{AB}^{(p,p)} \rho^{(p)} X_B^{(p)} + \frac{\rho^{(s)}}{\rho^{(p)}} \frac{\bar{n}}{X_A^{(p)}}} = X_B^{(p)} - \frac{\rho^{(s)}}{\rho^{(p)}} \bar{n} \tag{37}$$

To compare to simulations we will use the fraction of $p$ colloids bonded $k$ times $X_k^{(p)}$ which we obtain from Eqns. (20) - (22) as

$$\begin{aligned} X_o^{(p)} &= X_A^{(p)} X_B^{(p)} \\ X_1^{(p)} &= X_o^{(p)} \left( \frac{1}{X_A^{(p)}} + \frac{1}{X_B^{(p)}} - 2 \right) \\ X_2^{(p)} &= X_o^{(p)} \left( 1 - \frac{1}{X_A^{(p)}} \right) \left( 1 - \frac{1}{X_B^{(p)}} \right) \end{aligned} \tag{38}$$

We will place the $p$ colloids into two classes. The first class consists of $p$ colloids which are part of a chain which emanates from an $s$ colloid, we will call these star $p$ colloids with a colloid density $\rho_{arm}^{(p)}$. The second class consist of colloids which are not in a bonded network which includes an $s$ colloid, we will call these free $p$ colloids with a colloid density $\rho_{free}^{(p)}$ (note the monomer density $\rho_o^{(p)}$ is included in $\rho_{free}^{(p)}$). Unfortunately, the densities $\rho_{arm}^{(p)}$ and $\rho_{free}^{(p)}$ are not directly accessible with the current approach; however, the densities of free chains (including monomers) $\rho_{free}^{(chain)} = \sigma_B^{(p)}$ and star arms $\rho_{arm}^{(star)} = \rho^{(s)}\bar{n}$ are known. A quantity which will provide insight into the competition between self assembled star molecules and free chains is the fraction

$$\Psi = \frac{\rho^{(s)}\bar{n}}{\rho^{(s)}\bar{n} + \sigma_B^{(p)}} \tag{39}$$

Where $\Psi$ is the fraction of chains (free chains including monomers + star arms) which are star arms. The last quantity we would like to determine is the average length of the star arms $L_{arm}$. To determine this length (in a strict way) the densities $\rho_{arm}^{(p)}$ and $\rho_{free}^{(p)}$ would have to be known. They are not. As an alternative, an approximation of $L_{arm}$ can be constructed as follows. The total density of chains *not including* free monomers is $\rho_A^{(p)}$. For the star arms, the chains terminate on one side with a $p$ colloid (this colloid is bonded at both patches $A$ and $B$) bonded to an $s$ colloid and the other side with a $p$ colloid only bonded at patch $A$. To estimate $L_{arm}$ we will assume that the average number of double bonded colloids in a chain (not including free monomers in this average) is equal for free chains (not including monomers) and star arms. With this we can approximate $L_{arm}$ as follows

$$L_{arm} \approx \frac{\rho_{AB}^{(p)} + \rho_A^{(p)}}{\rho_A^{(p)}} = \frac{1}{X_B^{(p)}} \tag{40}$$

We will finalize this section with a discussion of the low $T$ limit of these quantities. Unlike a pure component two patch fluid which has the low $T$ limits $X_1^{(p)}\big|_{T\to 0} = 0$ and $X_2^{(p)}\big|_{T\to 0} = 1$, there will be, in general, a non – zero $X_1^{(p)}\big|_{T\to 0}$ in the current case of a mixture of $s$ colloids and two patch $p$ colloids. Using Eqns. (36) – (38), with $X_o^{(p)}\big|_{T\to 0} = 0$, we obtain the following limiting forms for $X_1^{(p)}$ and $X_2^{(p)}$

$$X_1^{(p)}\big|_{T\to 0} = X_B^{(p)}\big|_{T\to 0} = \frac{\rho^{(s)}}{\rho^{(p)}}\bar{n}\big|_{T\to 0} \tag{41}$$

$$X_2^{(p)}\big|_{T\to 0} = 1 - \frac{\rho^{(s)}}{\rho^{(p)}}\bar{n}\big|_{T\to 0}$$

Combining Eqns. (39) – (41) we obtain for $\Psi$ and $L_{arm}$

$$\Psi\big|_{T\to 0} = 1 \tag{42}$$

and

$$L_{arm}\big|_{T\to 0} = \frac{\rho^{(p)}}{\rho^{(s)}\bar{n}\big|_{T\to 0}} \tag{43}$$

These equations state that in the limit $T \to 0$ all $p$ colloids will be associated as star arms and there will be no free chains of $p$ colloids.

Equations (41) and (43) depend on the low $T$ limit of the average number of star arms per $s$ colloid $\bar{n}\big|_{T\to 0}$. Unfortunately, there does not appear to be a simple relation for this quantity. From Equations (19), (30) and (32) we obtain

$$\bar{n}\big|_{T\to 0} = \frac{\sum_{n=1}^{n^{max}} \frac{n}{n!}\left(\Delta\big|_{T\to 0}\right)^n \delta^{(n)} \Xi^{(n)}}{1+\sum_{n=1}^{n^{max}} \frac{1}{n!}\left(\Delta\big|_{T\to 0}\right)^n \delta^{(n)} \Xi^{(n)}} \qquad (44)$$

Now we solve for $\Delta\big|_{T\to 0}$ from Eqns. (36) and (41) as

$$\Delta\big|_{T\to 0} = \frac{\sqrt{\kappa_{AA}}}{4\pi d^2 (r_c - d)\kappa_{AB}}\left(\frac{\rho^{(p)}}{\rho^{(s)}\bar{n}\big|_{T\to 0}} - 1\right)\exp\left(\frac{(C-1)\varepsilon_{AB}^{(p,p)}}{k_B T}\right) \qquad (45)$$

Equations (44) – (45) provide a closed solution for $\bar{n}\big|_{T\to 0}$, which must be evaluated numerically. Once $\bar{n}\big|_{T\to 0}$ is obtained, the other low $T$ properties can be evaluated through the simple equations (41) – (43). For the case $C = 1$ the exponential in (45) becomes unity and there is no temperature dependence in the low $T$ limit. As will be shown, for this case, the theory reduces to a stable limiting result at temperatures which are not too low.

## IV.  Simulations

To provide a quantitative test of the new theory we perform new monte carlo simulations for the case discussed in section III where the *A* and *B* patches are located on opposite poles of the *p* colloids. We use the potential parameters $r_c = 1.1d$ and $\beta_c^{(A)} = \beta_c^{(B)} = 27°$ such that only single bonding of a *p* colloid will occur. Constant *NVT* (number of colloids, volume, temperature) simulations were performed using standard methodology.[36] Each *NVT* simulation was allowed to equilibrate for $10^8$ – $10^9$ trial moves and averages where taken for another $10^8$ – $10^9$ trial moves. A trial move consists of an attempted relocation of a *s* colloid or an attempted relocation and reorientation of a *p* colloid. For each simulation we used a total of *N* = 864 colloids. Constant *NPT* (number of colloids, pressure, temperature) simulations were performed in the same manner as the *NVT* simulations with the addition of an attempted volume change each *N* trial moves.

## V. Results

In this section we compare theory and simulation results for the case that the *p* colloid has two patches as discussed in sections III – IV and illustrated in Fig. 1.

### A. Dependence on composition

In this subsection we compare theory and simulation for the case that the association energy between patchy colloids is set to $\varepsilon^* = \varepsilon_{AB}^{(p,p)}/k_BT = 7$ with the association energy ratio in Eq. (33) set to $C = 1$. We consider both low density $\rho^* = \rho d^3 = 0.2$ and high density $\rho^* = 0.7$ cases. Figure 3 gives the fraction of *p* colloids bonded *k* times, Eq. (38), versus mole fraction of *s* colloids $x^{(s)}$. For $x^{(s)} \to 0$ the fluid is a pure component fluid of *p* colloids with $X_1^{(p)}$ being the dominant fraction for the low density case and $X_2^{(p)}$ being dominant for the high density case. Introducing *s* type colloids into the system, increasing $x^{(s)}$, results in a decrease in $X_2^{(p)}$ and $X_o^{(p)}$ with an increase in $X_1^{(p)}$. The decrease in $X_2^{(p)}$ and increase in $X_1^{(p)}$ is a result of the fact that longer free chains are being sacrificed to form star arms. Since $C = 1$ there is no energetic difference between a bond between two *p* colloids (*pp* bond) and a bond between *s* and *p* colloids (*sp* bond); however, the penalty in decreased orientational entropy for forming a *pp* bond is double that which is paid for an *sp* bond. Theory and simulation are in excellent agreement.

Figure 4 shows calculations for the average number of arms (bonds) per *s* colloid $\bar{n}$, fraction of chains which are star arms $\Psi$ and average length of star arms $L_{arm}$. For $x^{(s)} \to 0$ the *s* colloids are dilute. Since there are an abundance of *p* colloids available to "solvate" the *s* colloids, it is in this realm where $\bar{n}$ is a maximum and $\Psi$ is a minimum. It is also in this region where $L_{arm}$ is maximum. In fact, for $x^{(s)} \to 0$, $L_{arm} = L_{free}$ where $L_{free}$ is the average length of free chains. Increasing $x^{(s)}$ results in a decrease in $\bar{n}$ and $L_{arm}$ as there is now less *p* colloids to solvate the *s* colloids and the introduction of *s* colloids breaks longer chains of *p* colloids. As expected, increasing $x^{(s)}$ results in an increase in $\Psi$ as there are now more *s* colloids to seed star arms. As $x^{(s)} \to 1$, there are now few *p* colloids to solvate the *s* colloids and the probability of forming *pp* bonds becomes very small. In this limit $\bar{n} \to 0$, $\Psi \to 1$ and $L_{arm} \to 1$. According to these results, if one desired to create a small number of colloidal star molecules with many long arms, this is best achieved for small $x^{(s)}$. On the other hand, if one wished to create a larger number of stars with a few short arms, this would be best achieved for larger $x^{(s)}$. Comparing the cases for low and high density we see that increasing density increases $\bar{n}$, $\Psi$ and $L_{arm}$. Overall the theory and simulation are in excellent agreement; however, the theory predicts $\bar{n}$ to be too small for $x^{(s)} \to 0$ at high density. This is the result of the approximation of the *n* + 1 body cavity correlation functions as discussed in or previous paper.[23]

Figure 5 shows the excess internal energy $E^* = E_{AS}/Nk_BT$ and compressibility factor $Z = P/\rho k_BT$ for these same conditions. Initially, for $x^{(s)} \to 0$, increasing $x^{(s)}$ results in a decrease in both $E^*$ and Z as association in the system is increased. On the other extreme, $x^{(s)} \to 1$, *p* colloids are limiting and increasing $x^{(s)}$ decreases association in the system. This results in an increase in both $E^*$ and Z. Association is maximized at the location that $E^*$ and Z show distinct minimums, about $x^{(s)} \approx 0.1$ for $\rho^* = 0.7$ and $x^{(s)} \approx 0.21$ for $\rho^* = 0.2$. Theory and simulation are in excellent agreement for $E^*$. To provide a quantitative test for the theoretical predictions of Z we performed NPT simulation for various *s* colloid mole fractions $x^{(s)}$ as a function of $\rho^*$. These results can be found in Fig. 6. As can be seen, theory and simulation are in good agreement.

## B. Dependence on $\varepsilon^*$ at fixed composition with $C = 1$

Now we will analyze the effect of association energy $\varepsilon^*$ on fluid properties when composition remains constant. Specifically we will consider the case where $x^{(s)} = 0.05787$; again, we will keep the energy of a *pp* bond equal to a *ps* bond $\varepsilon_{AB}^{(p,p)} = \varepsilon_A^{(s,p)}$. We perform calculations for both low $\rho^* = 0.2$ and high $\rho^* = 0.7$ density cases. Figure 7 shows the fractions of *p* colloids bonded *k* times versus $\varepsilon^*$ for this case. For small $\varepsilon^*$, the monomer fraction $X_o^{(p)}$ is the dominant contribution, due to the fact that the entropic penalty of bond formation outweighs the energetic benefit for this low $\varepsilon^*$. Increasing $\varepsilon^*$ results in an increase in $X_1^{(p)}$ as the energetic benefit of forming a single bond outweighs the entropic penalty. At around $\varepsilon^* = 5$ for $\rho^* = 0.2$ and $\varepsilon^* = 3$ for $\rho^* = 0.7$ the fraction $X_2^{(p)}$ begins to have a significant increase with increasing $\varepsilon^*$. This increase in $X_2^{(p)}$ results in a maximum in $X_1^{(p)}$ near $\varepsilon^* = 7.5$ for $\rho^* = 0.2$ and $\varepsilon^* = 5.5$ for $\rho^* = 0.7$. Eventually the energetic benefit of being fully bonded outweighs the entropic penalty and $X_2^{(p)}$ becomes the dominant fraction. Overall theory and simulation are in excellent agreement, although for the density $\rho^* = 0.7$, the theory does slightly overpredict $X_2^{(p)}$ and underpredict $X_1^{(p)}$ for large $\varepsilon^*$.

Figure 8 shows calculations for the average number of arms (bonds) per *s* colloid $\bar{n}$, fraction of chains which are star arms $\Psi$ and average length of star arms $L_{arm}$; while Fig. 9 gives the results for the excess internal energy $E^* = E_{AS}/Nk_BT$. Increasing $\varepsilon^*$, or equivalently decreasing $T$, results in an increase in $\bar{n}$, $\Psi$, $L_{arm}$ and $|E^*|$ as association in the system increases. Overall theory and simulation are in excellent agreement for each quantity at low density. For the high density case theory and simulation are in good agreement, although the theory underpredicts $\bar{n}$ and overpredicts $L_{arm}$ for large $\varepsilon^*$. The simulation results for the fraction $\Psi$ seem to support the prediction given by Eq. (42), that at low $T$ there are few free chains and most $p$ colloids are associated into star arms.

### C. Dependence on association energy ratio $C$

Now we consider the specific effect of the ratio $C$ defined by Eq. (33). For $C < 1$ the $pp$ attractions are stronger than $ps$ attractions, while for $C > 1$ the opposite is true. Figure 10 shows calculations for the average number of arms (bonds) per $s$ colloid $\bar{n}$, fraction of chains which are star arms $\Psi$ and average length of star arms $L_{arm}$ versus $\varepsilon^*$ at a density of $\rho^* = 0.2$ and composition $x^{(s)} = 0.05$ for ratios $C = 0.6, 0.8, 1, 1.2$.

First we focus on the case C = 1.2. For this case the *sp* attractions are greater than *pp* attractions. Since *sp* attractions are favored, $\bar{n}$ is large, $\bar{n} \sim 7.3$ at $\varepsilon^* = 10$, and increases steadily as $\varepsilon^*$ is increased. For the studied cases, $\bar{n}$ is a maximum at C = 1.2. In contrast, the average star arm length $L_{arm}$ is a minimum for this case. This is a direct result of the large $\bar{n}$. Since there are a large number of arms and a finite number of $p$ colloids, the arms must be shortest for this case. Initially $L_{arm}$ increases with $\varepsilon^*$ as association increases in the system, goes through a maximum and then decreases as *pp* bonds are traded for *ps* bonds. We also note that $\Psi$ is a maximum for this case.

Now considering the case $C = 0.6$, where *pp* attractions are greater than *ps* attractions, we see that $\bar{n}$ initially increases with $\varepsilon^*$, goes through a maximum as *sp* bonds are traded for *pp* bonds and then decreases as $\varepsilon^* \to \infty$. It is for this $C$ that $\bar{n}$ is a minimum, $\bar{n} \sim 1$ at $\varepsilon^* \sim 10$, and $L_{arm}$ is a maximum, $L_{arm} \sim 4.5$ at $\varepsilon^* \sim 10$. Increasing $\varepsilon^*$ further to $\varepsilon^* \sim 20$ we find $\bar{n} \sim 0.34$ and $L_{arm} \sim 550$, meaning the *s* colloids exist primarily as monomers and chain ends for long chains of *p* colloids. For the case $C = 0.8$ we see similar behavior to the case $C = 0.6$, although less pronounced. Finally considering the case $C = 1$, we see that there are no maximums and both $L_{arm}$ and $\bar{n}$ reach a low temperature limiting value around $\varepsilon^* \sim 12$. From these results it is clear that the attraction ratio $C$ can be tuned to achieve a range of colloidal star molecules.

In the top panel of Fig. 10 we include the low $T$ limit of $\bar{n}$ obtained through Eqns. (44) – (45). For $C = 1$, 1.2 the low $T$ limit is reached near $\varepsilon^* \sim 12$ with the low $T$ limit giving reasonable predictions for $\varepsilon^* \geq 10$. When $C < 1$ the low $T$ limit is attained at a higher $\varepsilon^*$. It should be expected that the larger the ratio $C > 0$, the wider the range of applicability of the low $T$ limit.

### D. Low $T$ limit for the case $C = 1$

As a last case, we consider the low $T$ limits of the *p* colloid bonding fractions $X_1^{(p)}$ and $X_2^{(p)}$, average number of arms (bonds) per *s* colloid $\bar{n}$ and average length of star arms $L_{arm}$ as a function of *s* colloid mole fraction $x^{(s)}$. We use the limits developed in section III. These results are presented in Fig. 11. We consider the case that $\rho^* = 0.2$ and $C = 1$; however, the low $T$ results are very weakly dependent on density. In the limit of a pure fluid of *p* colloids we see $X_1^{(p)}\big|_{T\to 0} = 0$ and $X_2^{(p)}\big|_{T\to 0} = 1$. In this limit energy dominates and there is a single infinitely long chain of $p$ colloids. As we introduce $s$ colloids, increasing $x^{(s)}$, some of the $p$ colloids associate into star arms which breaks apart this infinitely long chain. This results in a decrease in $X_2^{(p)}\big|_{T\to 0}$ and increase in $X_1^{(p)}\big|_{T\to 0}$. This trend continues until $X_1^{(p)}\big|_{T\to 0}$ becomes unity and $X_2^{(p)}\big|_{T\to 0}$ vanishes as $x^{(s)} \to 1$ and interactions between $p$ colloids become rare. The average number of arms per $s$ colloid $\bar{n}\big|_{T\to 0}$ is at a maximum $\bar{n}\big|_{T\to 0} = n^{max} = 13$ when $x^{(s)} \to 0$ and all $p$ colloids can be associated into a single colloidal star. Initially, increasing $x^{(s)}$ results in a rapid decrease in $\bar{n}\big|_{T\to 0}$. For instance, increasing $x^{(s)}$ slightly to $x^{(s)} \sim 0.01$ results in a decrease of the average number of star arms per $s$ colloid to $\bar{n}\big|_{T\to 0} \sim 7.5$. This behavior is also seen in the term $L_{arm}\big|_{T\to 0}$ which is infinitely large for $x^{(s)} \to 0$ and decreases to $\sim 13$ at $x^{(s)} \sim 0.01$.

These limiting results provide guidance on the types of colloidal stars which will form for a given mole fraction $x^{(s)}$. Considering the results of Fig. 8, we see that this system had nearly reached it's low $T$ limit at the reduced temperature $T^* = 1/\varepsilon^* = 0.0833$, and the low $T$ limit provides a reasonable estimate of the average arm numbers and lengths for even higher temperatures.

Throughout this paper, for convenience of presentation, we have only considered the average number of bonds per $s$ colloid $\bar{n}$. However, the fraction of $s$ colloids bonded $n$ times (with $n$ star arms) $X_n^{(s)}$ is readily available through Eq. (19). We show the distribution of these fractions in the low $T$ limit, with $C = 1$, for three $s$ colloid mole fractions in Fig. 12. For the very dilute case $x^{(s)} = 10^{-5}$ the distribution is sharply peaked and asymmetric with the only two significant fractions being for $n = 11$ and $n = 12$. Increasing $s$ colloid mole fraction to $x^{(s)} = 0.02$ the distribution is symmetric and peaked at $n = 7$ with significant contributions between $n = 5$ and $n = 9$. Lastly, for the equimolar case $x^{(s)} = 0.5$, the distribution is highly asymmetric with the dominant contributions coming from $n = 0$ and 1 and significant contributions from $n = 2$ and $n = 3$. An interesting feature of this system is that even in the low $T$ limit where energy dominates there is still an entropic contribution which stems from the number of times a given $s$ colloid can bond.

## VI. Conclusions

We have developed a new theory to model binary mixtures of multi – patch *p* colloids and colloids with spherically symmetric attractions (*s* colloids). We developed the theory in Wertheim's [26, 32] multi – density formalism for associating fluids using modifications of the graphs developed in our previous paper[23] for the case of a mixture of single patch colloids and *s* colloids. We applied the theory to the case of a mixture of bi-functional *p* colloids, consisting of an *A* patch and *B* patch located on opposite poles of the colloid, and *s* colloids which only attract the *A* patch of the *p* colloid. There were only *AB* attractions between the patches on the *p* colloid, and there were no attractions between *s* colloids. This system was shown to self assemble into a mixture of free chains and colloidal star molecules. The average arm length of star molecules, ratio of free chains to star arms and average number of arms per colloidal star can be manipulated by varying density, temperature, composition and the ratio of association energies. The theory was shown to be accurate in comparison to monte carlo simulation data. We will consider the case II discussed in the introduction in a future paper.

In the development of the theory we assumed there were no attractions between *s* colloids which allowed us to write the theory as a perturbation to a hard sphere reference fluid. To include attraction between the *s* colloids one would need to use an appropriate reference system for the *s* colloids. Besides the definition of the reference system, the general results presented here would still be valid.

The perturbation theory developed in this paper gives a simple and accurate theory for the prediction of the self assembled structures and thermodynamic properties of mixtures of *s* and *p* colloids. However, there is no information about fluid structure of in the theory. For this we need to go to integral equation theories. Interestingly, the model presented in this work shares many similarities with the theory of highly asymmetric electrolyte solutions which are composed of large highly charged polyions and small counter ions with lower charge. Like the case considered in this work, the number of times the polyion (*s* colloids) can bond is unrestricted and the counter ions (*p* colloids) are restricted to bond a maximum of *k* times. Multi – density integral equation theories[37-39], which draw inspiration from Wertheim's integral equation theory solution[40, 41] of the Smith – Nezbeda[42] model of associating fluids, have proven to be accurate in modeling highly asymmetric solutions. Extension of these models such that the counterion (*p* colloid) has a patchy orientation dependent potential (instead of spherically symmetric as is the case for ionic interactions) with multiple patches will allow for the prediction of the structure of *s* and *p* colloid mixtures of the type studied in this paper.[43]

**Acknowledgments**

The financial support of The Robert A. Welch Foundation Grant No. C – 1241 is gratefully acknowledged. The authors would like to thank Y. V. Kalyuzhnyi for useful discussion.

**Appendix: Calculation of thermodynamic quantities**

In this appendix we calculate the chemical potential $\mu$, pressure $P$ and excess internal energy $E^{AS}$. The simplest way to calculate the chemical potential is to use the Euler – Lagrange equation supplied by Wertheim[26]

$$\frac{\mu^{(k)}}{k_B T} = \frac{\mu_{HS}^{(k)}}{k_B T} + \ln X_o^{(k)} - \frac{\partial \Delta c^{(o)}/V}{\partial \rho^{(k)}} \tag{A1}$$

where $\mu_{HS}^{(k)}$ is the hard sphere reference chemical potential for component $k$. From Eq. (A1) we obtain for the $s$ colloids

$$\frac{\mu^{(s)}}{k_B T} = \frac{\mu_{HS}^{(s)}}{k_B T} + \ln X_o^{(s)} - \frac{1}{2} \sum_{A \in \Gamma^{(p)}} \left(1 - X_A^{(p)}\right) \rho^{(p)} \frac{\partial \ln y_{HS}(d)}{\partial \rho^{(s)}}$$
$$- \frac{\bar{n}}{2} \rho^{(s)} \frac{\partial \ln y_{HS}(d)}{\partial \rho^{(s)}} - \sum_{n=1}^{n^{\max}} X_n^{(s)} \rho^{(s)} \frac{\partial \ln \delta^{(n)}}{\partial \rho^{(s)}} \tag{A2}$$

and for the $p$ colloids

$$\frac{\mu^{(p)}}{k_B T} = \frac{\mu_{HS}^{(p)}}{k_B T} + \sum_{A \in \Gamma^{(p)}} \ln X_A^{(p)} - \frac{1}{2} \sum_{A \in \Gamma^{(p)}} \left(1 - X_A^{(p)}\right) \rho^{(p)} \frac{\partial \ln y_{HS}(d)}{\partial \rho^{(p)}}$$
$$- \frac{\bar{n}}{2} \rho^{(s)} \frac{\partial \ln y_{HS}(d)}{\partial \rho^{(p)}} - \sum_{n=1}^{n^{\max}} X_n^{(s)} \rho^{(s)} \frac{\partial \ln \delta^{(n)}}{\partial \rho^{(p)}} \tag{A3}$$

With the chemical potentials known the pressure is easily calculated through the relation

$$P = \sum_j \mu^{(j)} \rho^{(j)} - A/V \tag{A4}$$

Lastly we obtain the excess internal energy as

$$\frac{E^{AS}}{N} = \frac{\partial}{\partial \beta}\left(\frac{\beta A^{AS}}{N}\right) = x^{(s)}\left(\frac{1}{X_o^{(s)}}\frac{\partial X_o^{(s)}}{\partial \beta} + \frac{1}{2}\frac{\partial \overline{n}}{\partial \beta}\right) + \left(1 - x^{(s)}\right)\sum_{A\in\Gamma^{(p)}} \frac{\partial X_A^{(p)}}{\partial \beta}\left(\frac{1}{X_A^{(p)}} - \frac{1}{2}\right) \quad (A5)$$

Specifically for the bi – functional 2 patch case considered in III with $C = 1$ we have the derivatives

$$\frac{\partial \ln X_A^{(p)}}{\partial \beta} = \frac{\frac{\partial \ln f_{AB}^{(p,p)}}{\partial \beta}\left(\gamma - \gamma^2 + \frac{x^{(s)}}{1-x^{(s)}}\left(\overline{n^2} - \overline{n}^2\right)\right)}{\frac{x^{(s)}}{1-x^{(s)}}\left(\overline{n} - \overline{n}^2 + \overline{n}^2\right) + \gamma^2 - 1} \quad (A6)$$

Where $\gamma = \xi \rho^{(p)} X_o^{(p)} f_{AB}^{(p,p)} \kappa_{AB}$ and $\overline{n^2} = \sum_{n=0}^{n^{\max}} n^2 X_n^{(s)}$. The remaining derivatives are then calculated as

$$\frac{\partial \ln X_B^{(p)}}{\partial \beta} = -\gamma \psi$$

$$\frac{\partial \ln X_o^{(s)}}{\partial \beta} = -\overline{n} \psi \quad (A7)$$

$$\frac{\partial \overline{n}}{\partial \beta} = \left(\overline{n^2} - \overline{n}^2\right)\psi$$

where $\psi = \left(\frac{\partial \ln f_{AB}^{(p,p)}}{\partial \beta} + \frac{\partial \ln X_A^{(p)}}{\partial \beta}\right)$.

**Figures:**

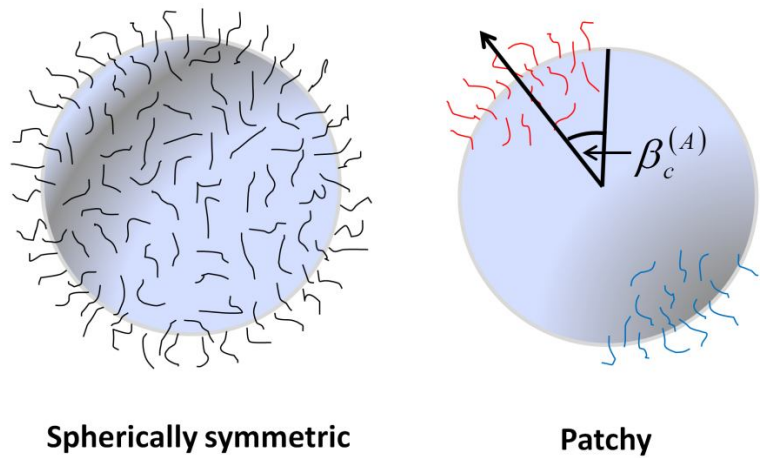

**Figure 1:** Illustration of spherically symmetric and two patch colloids

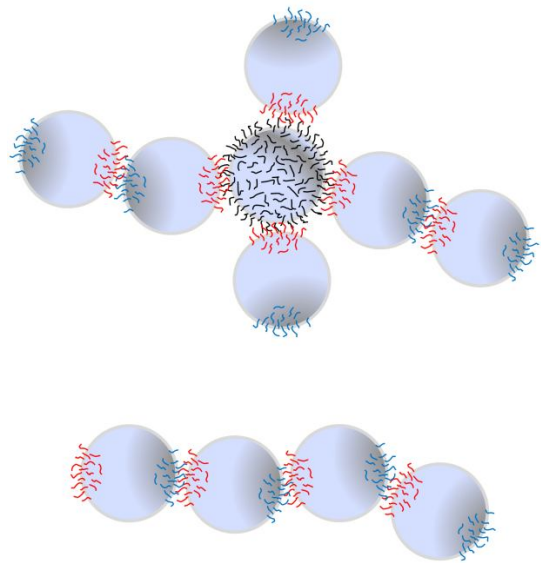

**Figure 2:** Examples of associated clusters which can be obtained from an *s* and *p* colloid mixture when the *p* colloids have two patches *A* and *B* and $\varepsilon_{AA}^{(p,p)} = \varepsilon_{BB}^{(p,p)} = \varepsilon_{B}^{(s,p)} = 0$

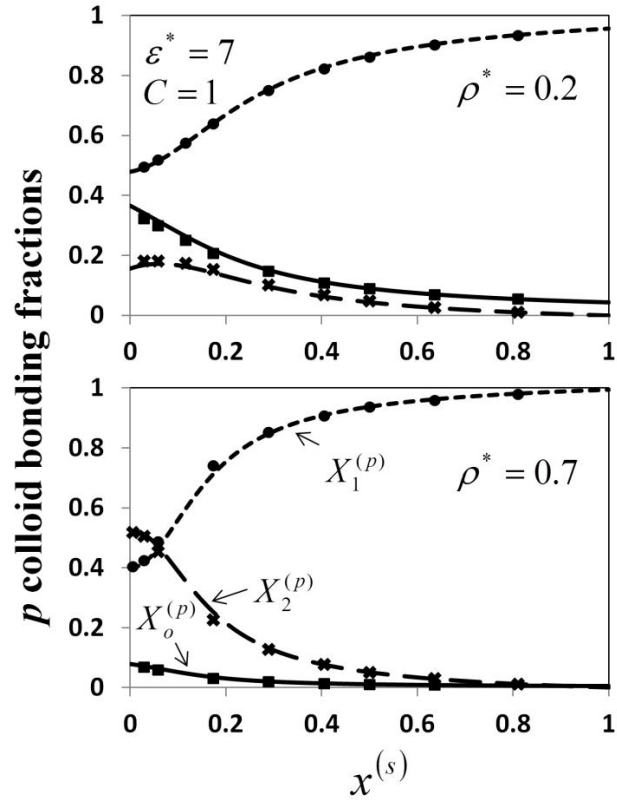

**Figure 3:** Fractions of $p$ colloids bonded $k$ times versus mole fraction of $s$ colloids $x^{(s)}$ at densities $\rho^* = 0.2$ (top) and $\rho^* = 0.7$ (bottom) and an association energy $\varepsilon^* = 1/T^* = \varepsilon_{AB}^{(p,p)}/k_B T = 7$ with $C = \varepsilon_A^{(s,p)}/\varepsilon_{AB}^{(p,p)} = 1$. Curves give theoretical predictions and symbols give simulation results: $X_o^{(p)}$ (solid curve – theory, squares – simulation), $X_1^{(p)}$ (short dashed curve – theory, circles – simulation), $X_2^{(p)}$ (long dashed curve – theory, crosses – simulation)

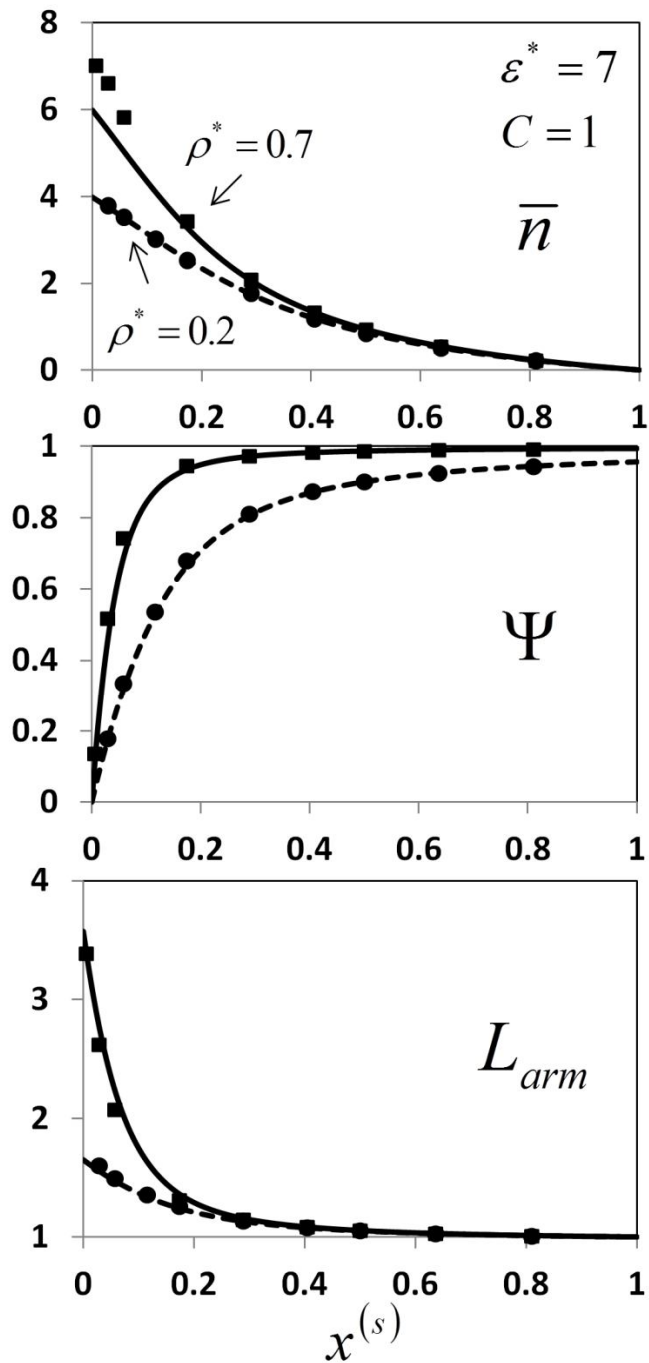

**Figure 4:** Average number of arms (bonds) per s colloid $\bar{n}$ (top), fraction of chains which are star arms $\Psi$ (middle) and average length of star arms $L_{arm}$ (bottom) versus mole fraction $s$ colloids $x^{(s)}$ at $\rho^* = 0.2$ (dashed curve – theory , circles – simulation) and $\rho^* = 0.7$ (solid curve – theory , squares – simulation). Association energy is $\varepsilon^* = \varepsilon_{AB}^{(p,p)}/k_B T = 7$ with $C = \varepsilon_A^{(s,p)}/\varepsilon_{AB}^{(p,p)} = 1$

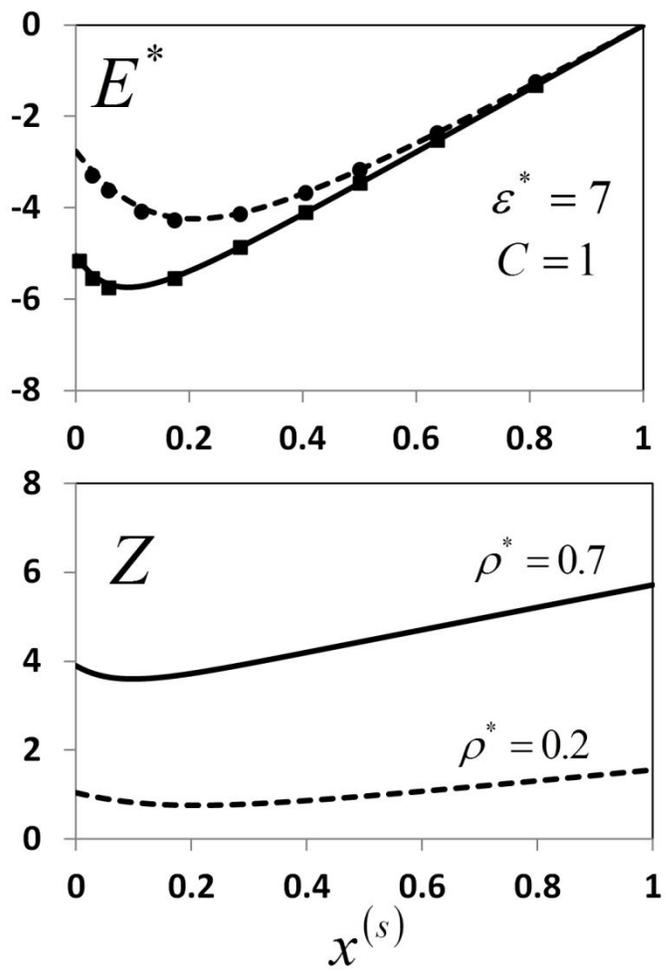

**Figure 5:** Same as Fig. 4 except the excess internal energy $E^* = E_{AS}/Nk_BT$ is the dependent variable (top) and compressibility factor $Z$ (bottom)

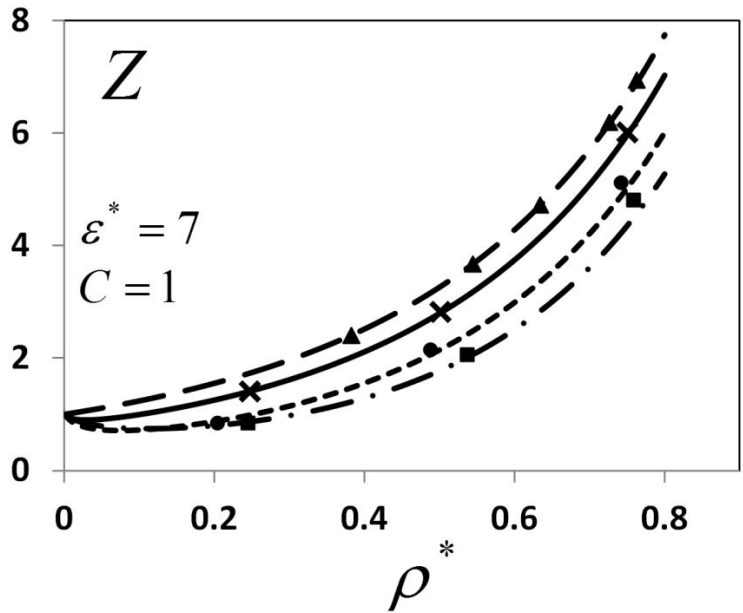

**Figure 6:** Compressibility factor Z versus density $\rho^*$ at an association energy $\varepsilon^* = \varepsilon_{AB}^{(p,p)}/k_B T = 7$ with $C = \varepsilon_A^{(s,p)}/\varepsilon_{AB}^{(p,p)} = 1$. Curves give theoretical predictions and symbols give *NPT* simulation results for $x^{(s)} = 1$ (long dashed curve – theory, triangles – simulation), $x^{(s)} = 0.752$ (solid curve – theory, crosses – simulation), $x^{(s)} = 0.405$ (short dashed curve – theory, circles – simulation) and $x^{(s)} = 0.116$ (dashed dotted curve – theory, squares – simulation)

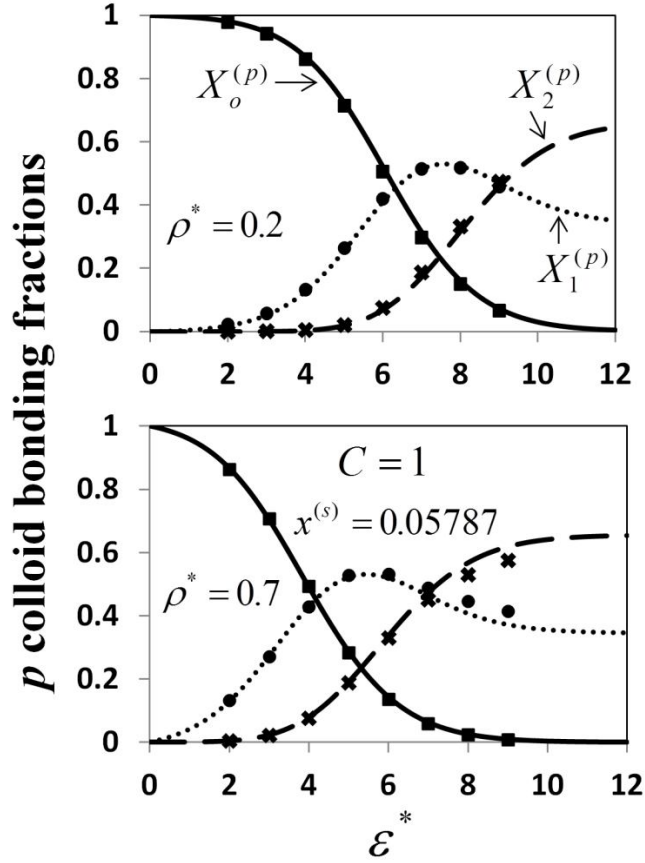

**Figure 7:** Fractions of $p$ colloids bonded $k$ times versus association energy $\varepsilon^* = \varepsilon_{AB}^{(p,p)}/k_BT$ at densities $\rho^* = 0.2$ (top) and $\rho^* = 0.7$ (bottom) and an $s$ colloid mole fraction $x^{(s)} = 0.05787$ with $C = \varepsilon_A^{(s,p)}/\varepsilon_{AB}^{(p,p)} = 1$. Curves give theoretical predictions and symbols give simulation results: $X_o^{(p)}$ (solid curve – theory, squares – simulation), $X_1^{(p)}$ (dotted curve – theory, circles – simulation), $X_2^{(p)}$ (long dashed curve – theory, crosses – simulation)

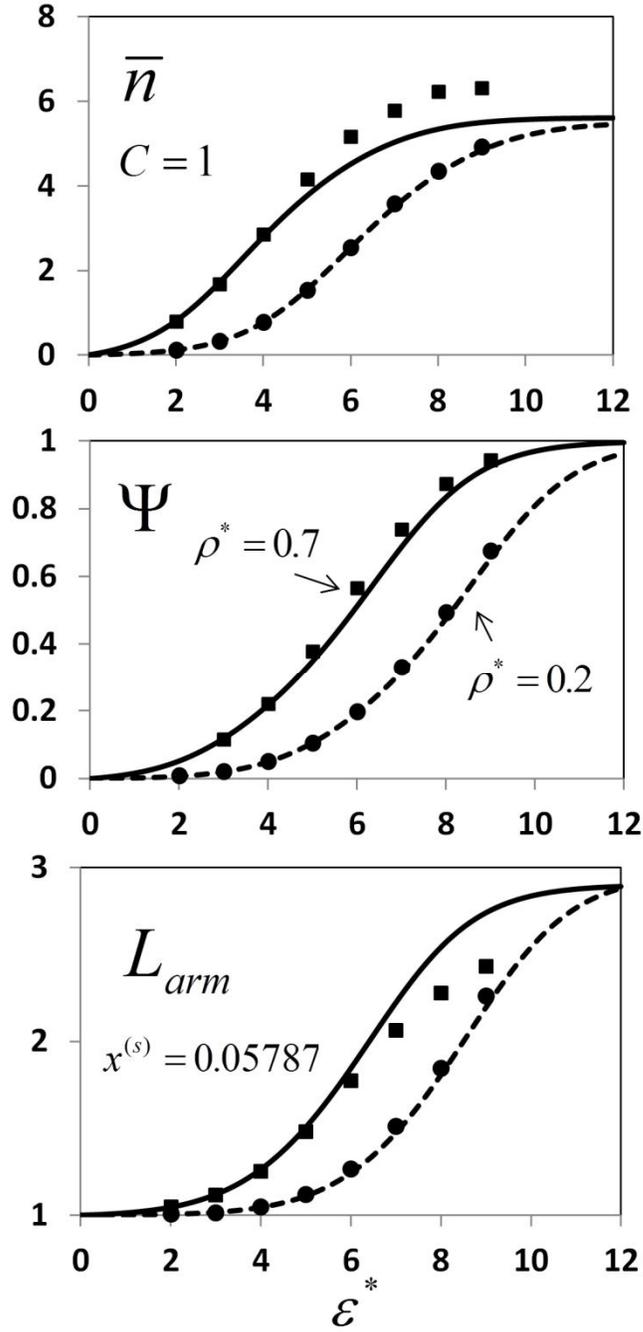

**Figure 8:** Average number of arms (bonds) per $s$ colloid $\bar{n}$ (top), fraction of chains which are star arms $\Psi$ (middle) and average length of star arms $L_{arm}$ (bottom) versus association energy $\varepsilon^*$ = $\varepsilon_{AB}^{(p,p)}/k_BT$ at $\rho^* = 0.2$ (dashed curve – theory , circles – simulation) and $\rho^* = 0.7$ (solid curve – theory , squares – simulation). Mole fraction of $s$ colloids is $x^{(s)} = 0.05787$ with $C = \varepsilon_A^{(s,p)}/\varepsilon_{AB}^{(p,p)} = 1$

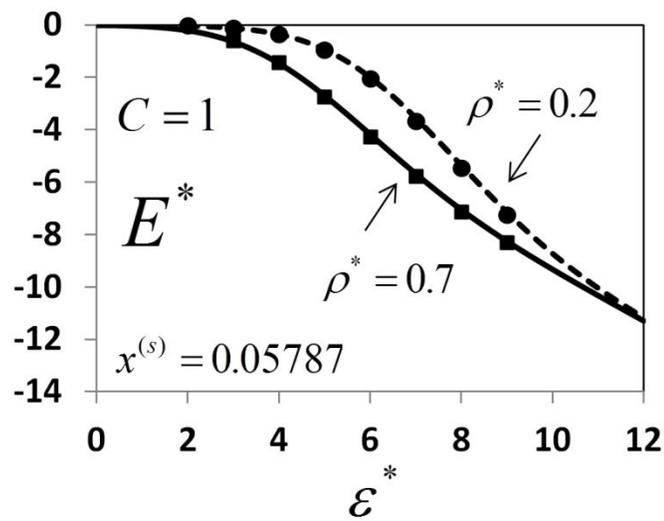

**Figure 9:** Same as Fig. 8 except the dependant variable is now excess internal energy $E^* = E_{AS}/Nk_BT$

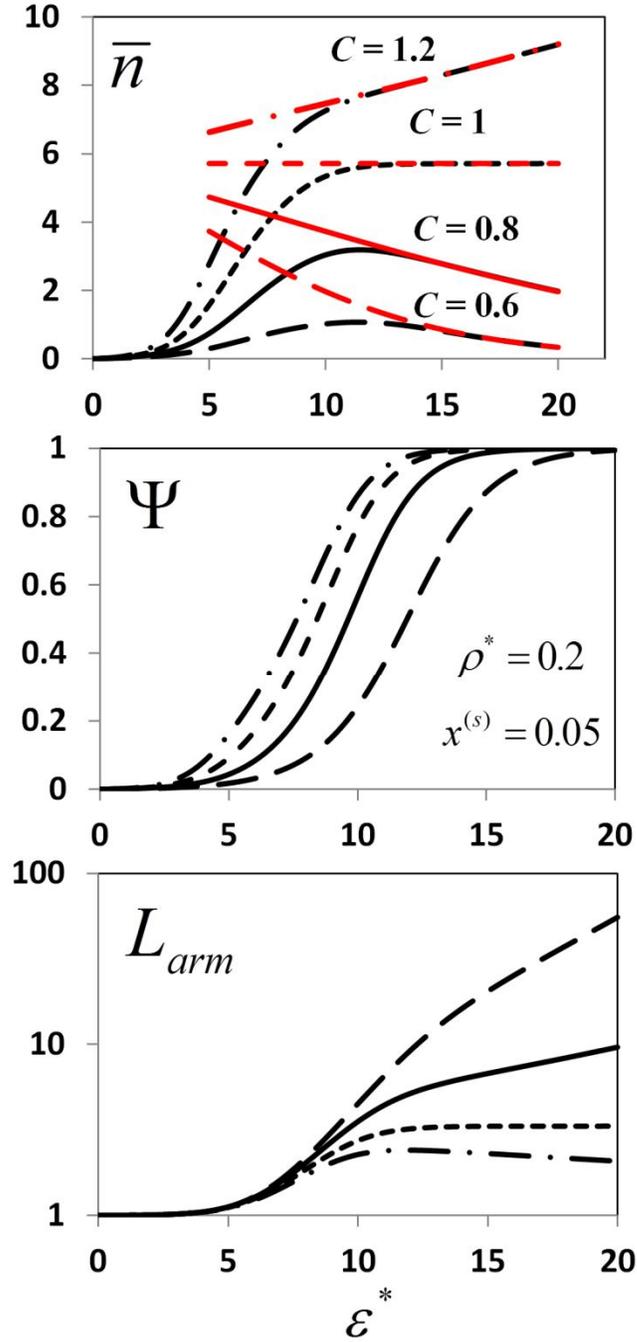

**Figure 10:** Average number of arms (bonds) per $s$ colloid $\bar{n}$ (top), fraction of chains which are star arms $\Psi$ (middle) and average length of star arms $L_{arm}$ (bottom) versus association energy $\varepsilon^* = \varepsilon_{AB}^{(p,p)}/k_B T$. Density is $\rho^* = 0.2$, with an $s$ colloid mole fraction $x^{(s)} = 0.05$ for various values of association energy between the $p$ and $s$ colloids $\varepsilon_A^{(s,p)} = C\varepsilon_{AB}^{(p,p)}$. Red curves beginning at $\varepsilon^* = 5$ on top panel give low T limit of $\bar{n}$ through Eqns. (44) – (45)

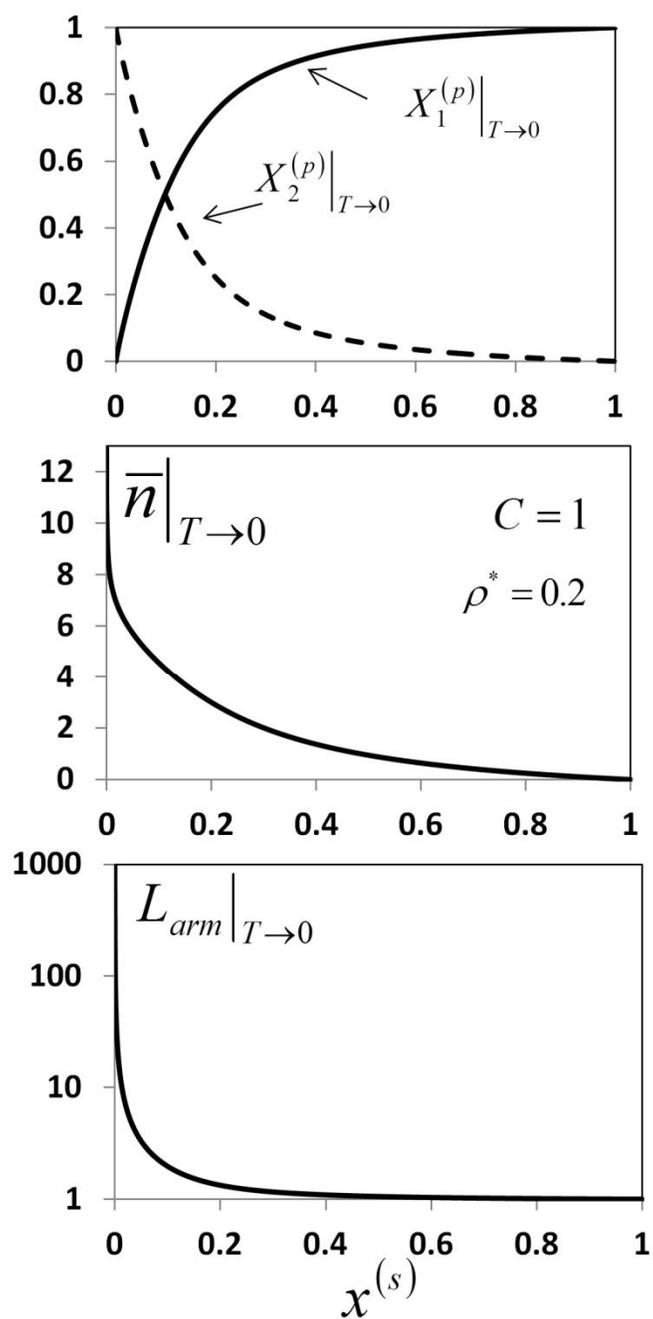

**Figure 11:** Low temperature limits of the fraction of *p* colloids bonded one or two times (top), average number of arms per *s* colloid (middle) and average arm length of star arms (bottom) versus mole fraction of *s* colloids. Association energy ratio is $C = 1$ and the density is fixed at $\rho^* = 0.2$

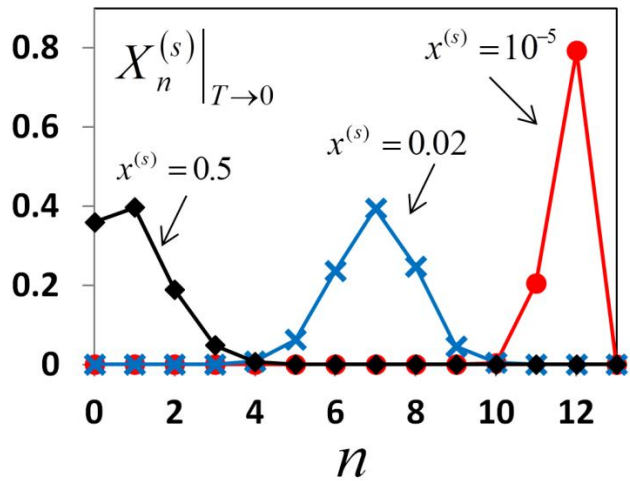

**Figure 12:** Low $T$ limit of fractions $X_n^{(s)}$ at a density of $\rho^* = 0.2$ and $C = \varepsilon_A^{(s,p)}/\varepsilon_{AB}^{(p,p)} = 1$. Symbols give theoretical predictions and lines are simply meant to guide the eye